\documentstyle[12pt]{article}
\title{The Equivalence Theorem and Global Anomalies}

\author{John F. Donoghue and Jusak Tandean\\[5mm]
Department of Physics and Astronomy\\
University of Massachusetts\\
Amherst, MA ~01003}
\begin{document}
\begin{titlepage}

\maketitle
\begin{abstract}

In the presence of some forms of global anomalies, the equivalence theorem,
which relates the interactions of longitudinal gauge bosons to those of the
Goldstone bosons, is not always valid.  This can occur when the Goldstone
sector contains an anomaly which is canceled in the gauge currents by the
effects of a different sector of the theory.  The example of the Standard Model
without Higgs particles is used to illustrate this phenomena.
\end{abstract}
{\vfill UMHEP-381}
\end{titlepage}

In the symmetry breaking of the Standard Model, three of the components of the
complex Higgs doublet become the longitudinal degrees of freedom of the
$W^\pm,Z^0$.  In alternative mechanisms for symmetry breaking, such as
Technicolor \cite{1,2}, there also exist spin zero particles which turn into
the
longitudinal components of the gauge bosons.  The equivalence theorem
\cite{3} says that the scattering amplitudes of longitudinal gauge bosons at
high
energy become equal to those of these spin zero Goldstone bosons in the
original theory, up to corrections suppressed by powers of the energy
\begin{equation}
 M(W_L^\pm, Z_L, \ldots) = M(w^\pm, z, \ldots) + O(M_W/E)
\end{equation}

\noindent Here $w^\pm,z$ are the Goldstone fields.  The equivalence theorem is
useful
because amplitudes involving the Goldstone bosons are generally easier to
analyze than the full gauge boson amplitudes.

In this paper we discuss fermionic theories of symmetry breaking in which there
are global but not gauge anomalies.  We will see that this can lead to some
counter examples to the equivalence theorem.  Roughly stated, some global
anomalies can modify the couplings of the Goldstone bosons, while not
changing the gauge boson couplings.  In most theories of Technicolor the
quantum numbers have been arranged in a way such that this does not occur,
but it remains a possibility in the larger framework of fermion-driven symmetry
breaking.

It is generally accepted that the quantum numbers of the fermion of the theory
must be chosen such that there are no anomalies in any of the currents coupled
to gauge bosons.  If the vector and axial vector gauge currents are described
by
matrices $T_V^{(a)}, T_A^{(a)}$
\begin{equation}
J_\mu^{(a)} = \bar{\psi }\gamma_\mu (T_V^{(a)} + T_A^{(a)} \gamma_5) \psi
\end{equation}

\noindent the anomaly-free requirement is that
\begin {equation}
D^{abc} = Tr(T_A^{(a)} \{T_V^{b}, T_V^{c}\}) = 0
\end{equation}

\noindent For example, with the $SU(2)_L$ part of the neutral weak current and
the
electromagnetic current for one family
\begin{eqnarray}
J^3_\mu & = & {1\over 2}[\bar{\gamma_\mu} (1 + \gamma_5)u - \bar{d} \gamma_\mu
(1 + \gamma_5)d + \bar{\nu} \gamma_\mu (1 + \gamma_5)\nu - \bar{e}
\gamma_\mu (1 + \gamma_5)e] \nonumber\\
J^\gamma_\mu & = & {2\over 3} \bar{u} \gamma_\mu u - {1\over 3} \bar{d}
\gamma_\mu d - \bar{e} \gamma_\mu e
\end{eqnarray}

\noindent the condition $D^{388} = 0$ occurs through the cancellation of the
quark and
electron contributions.  In a one-doublet model of Technicolor, where the
$SU(2)_L$ current is
\begin{equation}
J^{(3)}_\mu = {1\over 2} [\bar{U} \gamma_\mu (1 + \gamma_5) U - \bar{D}
\gamma_\mu (1 + \gamma_5)D]
\end{equation}
\noindent a cancellation with leptons is not used, so that the U(1) quantum
numbers must
be rearranged, i.e.,
\begin{equation}
J^\gamma_\mu = {1\over 2} [\bar{U} \gamma_\mu U - \bar{D} \gamma_\mu D]
\end{equation}

\noindent in order to generate the anomaly-free condition within a single
doublet.

In general there are other currents which do have anomalies.  For example, the
left handed isospin current in the Standard Model.
\begin{equation}
J^{3q}_\mu = {1\over 2}[\bar{u} \gamma_\mu(1 + \gamma_5)u - \bar{d}
\gamma_\mu(1 + \gamma_5)d],
\end{equation}

\noindent which is classically a global symmetry current in the limit that the
u, d quarks are
massless, does have an anomaly \cite{4}
\begin{equation}
\partial^\mu J_\mu = {\alpha N_c\over 12\pi} \epsilon^{\mu\nu\alpha\beta}
F_{\mu\nu} F_{\alpha\beta}+ {\rm quark \  mass  \  terms}.
\end{equation}

\noindent By standard methods, this leads to the matrix element for $\pi^0
\rightarrow
\gamma\gamma$ decay
\begin{equation}
M(\pi^0 \rightarrow \gamma\gamma) = {\alpha N_c\over 3\pi F\pi}
\epsilon_{1\mu} p_{1\nu} \epsilon_{2\alpha} p_{2\beta}
\end{equation}

\noindent which is in good agreement with the experimental value.

Fermionic theories of symmetry breaking make use of the fact that dynamical
symmetry breaking of a global invariance can also at the same time break the
underlying gauge symmetry.  A common pedagogical example is given by QCD
with massless u, d quarks\cite{1}, which has a global $SU(2)_L \times
SU(2)_R$ chiral symmetry.  This is dynamically broken to $SU(2)_V$, with an
order parameter
\begin {equation}
< 0 \mid \bar{u}_L u_R \mid 0 > = < 0 \mid \bar{d}_L d_R \mid 0 > \not= 0
\end {equation}

\noindent and with the pions, $\pi^\pm \pi^0$, being the Goldstone bosons.
However, the
vacuum condensates of Eq. 10 also break the $SU(2)_L \times U(1)$, gauge
invariance of the electroweak sector.  If the Standard Model were to contain no
Higgs bosons, these QCD interactions would provide the dynamical breaking of
$SU(2)_L \times U(1)$, with the pions being eaten to form the longitudinal
components of $W^\pm Z^0$, and masses given by
\begin{eqnarray}
M^2_W = {1\over 4} g^2_2 F^2_\pi = (30 MeV)^2 \nonumber\\
M^2_Z = M^2_W/cos^2 \theta_W
\end{eqnarray}

\noindent Technicolor theories are often said to be like QCD because they are
modeled
on this pattern of symmetry breaking, but with a larger mass scale $[F_\pi
\rightarrow v = 246 GeV]$ and generally with different particle and quantum
number assignments.

QCD can also provide a pedagogical example for the clash between the
equivalence theorem and global anomalies.  As mentioned above, there exist a
coupling of $\pi^0$ to two photons.  In the Higgsless Standard Model, is the
longitudinal Z coupling $Z^0 \rightarrow \gamma\gamma$ coupling equal to
that of $\pi^0 \rightarrow \gamma\gamma$, as stated by the equivalence
theorem?  Actually the transition of $Z^0_L \rightarrow \gamma\gamma$ is
forbidden for on-shell photons by Yang's theorem\cite{5}, However even for off-
shell photons the relevant diagram vanishes in the Higgless Standard Model.
The vertex would be generated by the triangle diagram of Fig. 1 with electrons
and u, d quarks in the loop.  The electrons and quarks cancel because their
quantum numbers have been arranged to yield an anomaly-free current.  In
contrast the $\pi^0 \rightarrow \gamma\gamma$ is related to the triangle
diagram with only u, d quarks in the loop.  The difference between these
implies
that $M(Z^0_L \rightarrow \gamma\gamma) \not= M(\pi^0 \rightarrow
\gamma\gamma)$.  However this is not as yet a true violation to the equivalence
theorem because it is not a process which occurs at high energies $E >>
M_W$.

In order to be convinced that the global anomalies can lead to a violation of
the
equivalence theorem, we have carried out the calculation for $e^+e^-
\rightarrow Z^0_L \gamma$ and compared it to $e^+e^- \rightarrow
\pi^0\gamma$ in the above model.  Both of these proceed through an off-shell
$\gamma$ (and Z) coupling.  We display only the intermediate photon result,
although the $Z^0$ contribution is very similar.  The $\pi^0\gamma$ final state
determined by the matrix element of Eq. 9, and yields
\begin{equation}
\sigma(e^+e^- \rightarrow \pi^0\gamma) = {\alpha^3\over 24\pi^2F^2_\pi}.
\end{equation}

\noindent The $Z^0\gamma$ amplitude with the quarks in the triangle diagram
requires
the full loop amplitude which has been given by Adler.  For massless quarks
one finds
\begin{equation}
\sigma(e^+e^-  Z\gamma)_{quark} =  {\alpha^2 g^2_2\over 96\pi^2 M^2_Z cos^2
\theta_w} \left(1 - {M^4_Z \over q^4} \right) \left(1 - {M^2_Z \over q^2 -
M^2_Z}
\right)^2
\end{equation}

\noindent Because the vector boson mass is related to $F_\pi$ as given in Eq.
11, Eq. 13
would satisfy the equivalence theorem if only quarks were to be included.
However the full calculation of the $Z\gamma$ final state requires both quarks
and leptons in the triangle diagram, and the lepton couplings have been
arranged to cancel the effects of quarks, such that if all the fermions are
massless one finds
\begin{equation}
\sigma(e^+e^- \rightarrow Z^0_L \gamma)_{TOT} = 0
\end{equation}

\noindent We have not calculated any higher order diagrams leading to the
$Z\gamma$
final state.

None of the existing discussions\cite{3} of the equivalence theorem take into
account the possibilities of global anomalies in the Goldstone boson sector.
Indeed most proofs are firmly within the context of the Standard Model with
Higgs particles, and the Higgs particles do not have any anomalous couplings.
In order to incorporate global anomalies into the treatment of longitudinal
gauge
bosons, one may use effective Lagrangians.  Consider a theory beyond the
Standard Model which has two sectors.  One is strongly interacting and
contains the fields which generate the Goldstone bosons.  The only role of the
other sector is to cancel the anomalies in the gauge currents, and we will
assume that it consists of weakly interacting particles.  If the
scale of the strong sector is well above $M_W$, its effect at low energy can be
described by an effective Lagrangian involving only the Goldstone fields.  The
result is rather similar to the effective Lagrangian found when one integrates
out
a very heavy fermion\cite{6}.  In that case, the heavy fermion does not
completely decouple, but leaves behind the effect of the anomaly.  Likewise, in
our example the low energy theory must contain an explicit
Wess-Zumino-Witten\cite{7} anomaly Lagrangian
in order to represent the contribution to the
anomaly of the heavy particles.  We have
\begin{eqnarray}
{\cal L}_{TOT} = L_{Goldstone} + L_{lepton}\nonumber\\
{\cal L}_{Goldstone} = L_{reg} + L_{anomaly}
\end{eqnarray}

\noindent Here $L_{lepton}$ is the usual $SU(2)_L \times U(1)$ invariant
Lagranging of the
weakly interacting sector.  The usual Goldstone boson effective Lagrangian is
\begin{eqnarray}
{\cal L}_{reg} & = & {v\over 4} Tr(D_\mu U D^\mu U^{\dag} ) + \ldots\nonumber\\
U & = & exp \left(i {{\vec {\tau}} \cdot {\vec {\phi}} \over
v}\right)\nonumber\\
D_\mu U & = & \partial_\mu U - i g_2 {{\vec {\tau}}\over 2}\cdot
{\vec {W}}_\mu U + ig_1 U{\tau_3 \over 2} B_\mu
\end{eqnarray}

\noindent The Wess Zumino Witten anomaly Lagrangian depends on the quantum
number assignments of the fundamental fields.  For a weak doublet in an
$SU(N)$ vector theory, the complete anomaly Lagrangian is given in Ref. 8.
We display here the portion involving two gauge bosons
\begin{eqnarray}
& &L_{anomaly} = -i {N_{TC}\over 48\pi^2} \epsilon^{\mu\nu\alpha\beta}
Tr[ \partial_\mu r_\nu U^{\dag} \ell_\alpha U R_\beta + \partial_\mu \ell_\nu U
r_\alpha U^{\dag} L_\beta\nonumber\\
& &+ (r_\mu \partial_\nu r_\alpha + \partial_\mu r_\nu r_\alpha)R_\beta +
(\ell_\mu
\partial_\nu \ell_\alpha + \partial_\nu \ell_\alpha + \partial_\mu \ell_\nu
\ell_\alpha)L_\beta\nonumber\\
& &+ {1\over 2} (r_\mu R_\nu r_\alpha R_\beta - \ell_\mu L_\nu \ell_\alpha
L_\beta) -
r_\mu U^{\dag} \ell_\nu U R_\alpha R_\beta + \ell_\mu U r_\nu U^{\dag} L_\alpha
L_\beta]
\end{eqnarray}

\noindent where
\begin{eqnarray}
L_\mu & = & \partial_\mu U U^{\dag} , R_\mu = U^{\dag} L_\mu U = U^{\dag}
\partial_\mu
U\nonumber\\
-i \ell_\mu & = & g_1 (Q - {\tau_3\over 2})B_{\mu} + g_2
{{\vec {\tau}} \over 2}\cdot {\vec {W}}_\mu \nonumber\\
&  = & eA_\mu Q + {g_2 Z_\mu\over cos\theta_w} [{\tau_3\over 2} -
sin^2\theta_wQ] + {g_2\over \sqrt 2} (\tau_+ W^-_\mu + \tau_-
W^+_\mu)\nonumber\\
-i r_\mu & = & g_1 Q B_\mu = e Q A_\mu - {g_2 Z_\mu\over cos \theta_w} sin^2
\theta_w Q
\end{eqnarray}

Given this effective Lagrangian, one can make a gauge change which removes
most of the manifestations of the Goldstone bosons.  For
this ``unitary gauge'' we
transform
\begin{equation}
g_2 {{\vec {\tau}}\over 2} \cdot {\vec {W}}_\mu
\rightarrow g_2 {{\vec {\tau}}\over 2} \cdot {\vec
{W}}^\prime_\mu = U g_2 {{\vec  {\tau}}\over 2} \cdot {\vec {W}}_\mu U^{\dag}
-i\partial_\mu U U^{\dag}
\end{equation}

\noindent The lowest order Lagrangian becomes
\begin{equation}
L_2 =  {g^2_2 F^2_\pi\over 4} (W^+_\mu W^{-\mu} + {g^2_2 F^2_\pi\over 8
cos^2\theta_w} A_\mu A^\mu)
\end{equation}

\noindent i.e., only the gauge boson mass term survives.  However the Goldstone
fields
do not disappear from the anomaly.  While the specific form is not instructive,
we
have verified that the anomaly Lagrangian still contains both $U$ and
$\partial_\mu U$ after the gauge transformation of Eq. 18.  Thus one does not
entirely remove the Goldstone degree of freedom by going to the unitary gauge.
This result is an indication, within the framework of effective Lagrangians, of
the
inequivalence of some of the Goldstone and gauge boson couplings

Most models of Techicolor do not share this problem.  The reason is that in
constructing a new technicolor model one generally requires that anomaly
cancellations occur completely within the new strongly interacting sector as in
Eq. 6.  To do otherwise would be uneconomical because one would require
further fermions (``leptons'') outside of the Technicolor sector in order to
make
the theory free of gauge anomalies.  However there is no requirement that
forbids such an arrangement of fermions, as can be seen from the fact that
Nature has chosen these quantum number assignments for the quarks and
leptons of the Standard Model.  Of course, the QCD effects described above
have negligible effects on W and Z physics at the TeV scale.  There may,
however, be TeV scale theories in the class of possible examples of fermionic
symmetry breaking which do contain global anomalies for the Goldstone
bosons.

{\bf Acknowledgements}: We thank A. Manohar, H. Georgi and G. Valencia for
useful discussions.
This work was supported in part by the U.S. National Science Foundation.

\end{document}